\documentclass[a4paper,11pt]{article}

\begin{document}

\newcommand{\w}{\omega}
\newcommand{\T}{\theta}
\newcommand{\qbar}{\overline{q}}
\newcommand{\e}{\mathrm{e}}
\newcommand{\rv}{r^{\vee}}
\newcommand{\hv}{h^{\vee}}
\newcommand{\bref}[1]{(\ref{#1})}
\newcommand{\be}{\begin{equation}}
\newcommand{\ee}{\end{equation}}
\newcommand{\ba}{\begin{eqnarray}}
\newcommand{\ea}{\end{eqnarray}}
\newcommand{\se}[1]{\section{#1}}
\newcommand{\rname}[1]{\newblock {\em #1.}}
\renewcommand{\thefootnote}{\fnsymbol{footnote}}

\hyphenation{Lagrangian}

\begin{flushright}
DTP-99/55 \\
\texttt{hep-th/9908184} \\
August 1999 \\
\end{flushright}

\vskip 0.9cm
\begin{center}
{\Large{\bf S-Matrix Identities in Affine Toda Field Theories}}
\end{center}
\vskip 0.4cm
\centerline{Peter Mattsson%
\footnote{E-mail: {\tt P.A.Mattsson@durham.ac.uk}}
}
\vskip 0.25cm
\centerline{\sl\small Dept.~of Mathematical Sciences,
University of Durham, Durham DH1 3LE, UK\,}
\vskip 0.9cm

\begin{abstract} 
We note that S-matrix/conserved charge identities in affine Toda field
theories of the type recently noted by Khastgir can be put on a more
systematic footing.  This makes use of a result first found by Ravanini,
Tateo and Valleriani for theories based on the simply-laced Lie algebras
(A, D and E) which we extend to the nonsimply-laced case. We also present
the generalisation to nonsimply-laced cases of the observation - for
simply-laced situations - that the conserved charges form components of
the eigenvectors of the Cartan matrix. 
\end{abstract}

\se{Introduction}
Affine Toda field theory (ATFT) is a massive scalar field theory in 1+1
dimensions, with a Lagrangian of the form
\be
\mathcal{L}=\frac{1}{2}\partial^{\mu}\phi_{a}\partial_{\mu}\phi_{a}
-\frac{m^{2}}{\beta^{2}}\sum_{j=0}^{r}n_{j}\exp(\beta\alpha_{j}\cdot \phi).
\ee
The $\alpha_{i}$, $j=1,\ldots,r$ are the simple roots of a rank-$r$
semi-simple Lie algebra $g$, $\alpha_{0}$ being the affine root. There
exists an ATFT related to each possible (twisted or untwisted) Dynkin
diagram; quantum scattering matrices (S-matrices) were first found for the
simply-laced cases~\cite{Arinshtein, Braden, Christe}, and later for
nonsimply-laced cases~\cite{Delius, Corrigan}. For theories based on
nonself-dual algebras, the algebras fall naturally into dual pairs
$(b_{n}^{(1)},a_{2n-1}^{(2)}), (c_{n}^{(1)},d_{n+1}^{(2)}),
(g_{2}^{(1)},d_{4}^{(3)})$, and $(f_{4}^{(1)},e_{6}^{(2)})$. The S-matrix
for the second member of each pair is related to that of the first by the
strong/weak coupling duality $\beta \rightarrow \frac{4\pi}{\beta}$. The
simply-laced algebras and $a_{2n}^{(2)}$ are self-dual.
  
In a recent paper, Khastgir~\cite{Khastgir} noted that there existed
product identities on elements of these S-matrices, all evaluated at the
same rapidity, associated with sum rules for the conserved charges of the
theory. For example, he noted that in $a_{r}^{(1)}, d_{r}^{(1)},
(c_{r}^{(1)},d_{r+1}^{(2)})$ and $(b_{r}^{(1)},a_{2r-1}^{(2)})$ we had
\ba
S_{22}(\T)=S_{11}(\T)S_{13}(\T),& &q_{s}^{2}q_{s}^{2}=q_{s}^{1}q_{s}^{1}
+q_{s}^{1}q_{s}^{3},
\label{eq:kexamp}
\ea
where $S_{ab}(\T)$ is an element of the S-matrix evaluated at rapidity
$\T$, and $q_{s}^{x}$ is the $x$th conserved charge at spin $s$. The above
is true for any $s$, with non-trivial independent conserved charges
existing for $s$ an exponent of the Lie algebra. This is a consequence of
the fact - first noted in~\cite{Braden} (see also~\cite{Klassen2}) - that
the coefficients of a Fourier series expansion of the logarithmic
derivative of the S-matrix provide solutions to the conserved-charge
bootstrap. If we define
\be
-i\frac{d}{d\T}\ln S_{ab}(\T)=\varphi_{ab}(\T)=-\sum_{k=1}^{\infty}
\varphi_{ab}^{(k)}\e^{-k|\T|},
\ee
then the linearly independent rows and columns of $\varphi^{(k)}$
each satisfy the conserved charge bootstrap, and so are proportional to
the unique vector $q_{k}$, implying $\varphi_{ab}^{(k)} \propto
q_{k}^{a}q_{k}^{b}$. This does allow the possibility that some of the
matrices $\varphi^{(k)}$ are zero, making their relationship to the
conserved charges trivial, but we shall see later that this does not
happen, at least for generic couplings. 

Khastgir did not give a general prescription as to how to find such
S-matrix identities, beyond noting that some can be written down by
inspecting the product-form S-matrix.  We give such a prescription, and
also expand the list of identities to cases when not all the S-matrices
are evaluated at the same rapidity.

In order to find these identities, we need to make use of an identity
found by Ravanini, Tateo and Valleriani (RTV)~\cite{Ravanini} for
simply-laced cases, and to generalise it to nonsimply-laced cases. We
establish this fundamental identity in Section~\ref{sec:basic} before
using it, in Section~\ref{sec:multi}, to generate further S-matrix
identities. 

\section{The Basic Identities}
\label{sec:basic}
In simply-laced cases, Ravanini, Tateo and Valleriani~\cite{Ravanini}
found
\be
S_{ab}\left(\T+\frac{i\pi}{h}\right)S_{ab}\left(\T-\frac{i\pi}{h}\right)=
\e^{-2i\pi G_{ab}\Theta(\T)}\prod_{c=1}^{r}S_{ac}(\T)^{G_{bc}},
\label{eq:rtv}
\ee
for the minimal S-matrix and gave a general proof based on the geometric
formulae of~\cite{Dorey}. This identity also holds for the full S-matrix.
Here, $G$ is the incidence matrix of the Dynkin diagram of the Lie
algebra, $h$ is the Coxeter number, and $\Theta$ is the step function
\be
\Theta(x)=\lim_{\epsilon \rightarrow 0}\left[\frac{1}{2}+\frac{1}{\pi}
\arctan \frac{x}{\epsilon}\right]=\left\{
\begin{array}{ccl}
0&\mathrm{if}&x<0, \\
\frac{1}{2}&\mathrm{if}&x=0, \\
1&\mathrm{if}&x>0.
\end{array} \right.
\ee

There also exist integral formulae for the S-matrix, given by
Oota~\cite{Oota} and Frenkel and Reshetikhin \cite{Frenkel}; these can
be used to prove RTV's formula but, since they apply to \emph{all}
simple Lie algebras, can also be used to generalise~\bref{eq:rtv} to
nonsimply-laced cases.

Before we can present Oota's formula, we must break off to provide a few
definitions. We shall work with the untwisted algebras here, since the
S-matrices for the theories based on twisted algebras can be found through
the duality transformation. (The exceptional case $a_{2n}^{(2)}$ will be
discussed at the end of this section.) Firstly, let $\hv$ be the dual
Coxeter number, $\rv$ be the maximum number of edges connecting any two
vertices of the Dynkin diagram\footnote{This is 1 for the self-dual 
cases, and 2 for the nonself-dual ones, except for 
$(g_{2}^{(1)},d_{4}^{(3)})$ where it is 3.}, and $0 \leq B \leq 2$ be a
function of the coupling constant, conjectured to be~\cite{Arinshtein,
Braden, Christe, Dorey2}
\be
B^{g}=\frac{2\beta^{2}}{\beta^{2}+\frac{4\pi h}{\hv}}
\ee 
for Lie algebra $g$. Next, let the set of simple roots of the Lie
algebra be $\{\alpha_{i}\}$,
and define $t_{i}=\rv\frac{(\alpha_{i},\alpha_{i})}{2}$, where the length
of the long roots is normalised to 2\footnote{Thus $t_{i}=1$ for short
roots and $t_{i}=\rv$ for long roots.}. Finally, define
\be
[K_{ij}]_{q\qbar}=(q\qbar^{t_{i}}+q^{-1}\qbar^{-t_{i}})\delta_{ij}
-[G_{ij}]_{\qbar},
\label{eq:kdef}
\ee
and
\be
M_{ij}(q,\qbar)=\left([K]_{q\qbar}\right)_{ij}^{-1}[t_{j}]_{\qbar},
\ee
where $q(t)=\exp \left(\frac{(2-B)t}{2h}\right)$, $\qbar(t)=\exp
\left(\frac{Bt}{2\rv\hv}\right)$ and we use the standard notation
$[n]_{q}=(q^{n}-q^{-n})/(q-q^{-1})$. Oota found that
\be
S_{ab}(\T)=(-1)^{\delta_{ab}}\exp\left(4\int_{-\infty}^{\infty}\frac{dk}{k}
\e^{ik\T}\left\{\sin k\T_{h}\cdot\sin k\T_{H}\cdot 
M_{ab}(q(\pi k),\qbar(\pi k)) +\frac{\delta_{ab}}{4}\right\}\right),
\label{eq:Oota}
\ee
where, for conciseness, we have defined $\T_{h}=\frac{i\pi(2-B)}{2h}$ and
$\T_{H}=\frac{i\pi B}{2\rv\hv}$. The formula given by Frenkel and
Reshetikhin~\cite{Frenkel} is similar to this, but without the factor of
$(-1)^{\delta_{ab}}\exp \left(\int_{-\infty}^{\infty}\frac{dk}{k}
\e^{ik\T}\delta_{ab}\right)$. The standard Fourier transform result
$\int_{-\infty}^{\infty}\frac{dk}{k}\e^{ik\T}=i\pi \mathrm{sgn\ } \T$,
together with the $2i\pi$ periodicity of the exponential, shows that this
factor is 1 for $\T$ real, though it is different from 1 if $\T$ is
complex. We shall use Frenkel and Reshetikhin's form here. 

We now propose a generalised RTV identity of the form
\be
S_{ij}(\T+\T_{h}+t_{i}\T_{H})S_{ij}(\T-\T_{h}-t_{i}\T_{H})=\e^{y}
\prod_{l=1}^{r}\prod_{n=1}^{G_{il}}S_{jl}(\T+(2n-1-G_{il})\T_{H}),
\label{eq:grtv}
\ee
and aim to find $y$. We now need to be careful, particularly for
$\T=0$, as in this case we can either consider the lhs as
\be
\lim_{X\rightarrow \T_{h}+t_{i}\T_{H}}\lim_{\T\rightarrow 0}
S_{ij}(\T+X)S_{ij}(\T-X)
\label{eq:limits}
\ee 
or with the limits reversed. In the first case (the one we would
choose if we were to go on and use this in the thermodynamic Bethe
ansatz approach) it simply becomes the unitarity constraint on S, and
thus is equal to 1. If we take the limits in
the other order, then (as can be seen if the S-matrix is written out in
block form) the result becomes -1 if the S-matrix has a pole at
$\T_{h}+t_{i}\T_{H}$. Similar situations can also arise for the rhs,
and in any other case where we have a product $S(\T+X)S(\T-X)$ such
that one term is evaluated at a pole of the S-matrix, and the other is
at a zero. 

Following RTV, we will consider the first case here. We found it simplest
to replace $\T_{h}$ and $\T_{H}$ in~\bref{eq:grtv} by $\T_{h}+i\epsilon$
and $\T_{H}+i\epsilon$, and take the limit $\epsilon \rightarrow 0$ last.
Substituting in \bref{eq:Oota} and simplifying, we find
\ba
\lefteqn{\e^{y}= \lim_{\epsilon \rightarrow 0}\exp
\left(\sum_{l=1}^{r}\int_{-\infty}^{\infty}\frac{dk}{k}\e^{ik\T}
\cdot \right.} \nonumber \\
&& \left. \left\{q(\pi
k)-q(-\pi k)\right\}\left\{\qbar(\pi k)-\qbar(-\pi k)
\right\}[K_{il}]_{q'(\pi k)\qbar'(\pi k)}M_{lj}(q(\pi k),\qbar(\pi k))\right)
\label{eq:ey}
\ea
where $q'(t)=q(t)\e^{\frac{\epsilon t}{\pi}}$ and
$\qbar'(t)=\qbar(t)\e^{\frac{\epsilon t}{\pi}}$. Looking ahead to
Section~\ref{sec:other} and formula~\bref{eq:mform}, we can see that when
the integrand in~\bref{eq:ey} is expanded out, all the terms are of the
form $t(x,\T)=\int_{-\infty}^{\infty}\frac{dk}{k}\e^{ik\T}\e^{x|k|}$, with
$x$ real, which is divergent if $x$ is positive. It is, however, implicit
in Oota's formulation that any terms which are na\"{\i}vely divergent must
be analytically continued. For $x$ negative, $t(x,\T)$ is just another
standard Fourier transform, which has the result $-2i \arctan
\frac{\T}{x}$, so, after analytic continuation, we should set $t(x,\T)=-2i
\arctan \frac{\T}{x}$ for all $x$.

It is helpful to split the $t(x,\T)$ terms into two sets: $x \rightarrow
0$ in the limit $\epsilon \rightarrow 0$ (but non-zero otherwise) and the
rest. It is only for the first of these that the choice of limit
prescription makes a difference. Taking the $\epsilon$ limit first, they
are $\pm i\pi$, depending on whether $x \rightarrow 0\mp$, for all $\T$,
but, with the limits taken the other way, we get $\lim_{x \rightarrow 0}
-2i\arctan \frac{\T}{x}$. 

If we had chosen to take the $\epsilon$ - rather than the $\T$ - limit
first, $\sum_{l=1}^{r}[K_{il}]_{q\qbar}M_{lj}(q(\pi k),\qbar(\pi k))$
would have reduced to $\delta_{ij}[t_{j}]_{\qbar(\pi k)}$. Each $t(x,\T)$
would then be matched by a $t(-x,\T)$, so the rhs would reduce to 1, i.e.
$y=0$. Looking carefully at the above reasoning, we can convert this to a
result about the other limit prescription by adding $-2i\pi \Theta(\T)$ to
$y$ for each term of the first type present, this being the difference
between its value with the $\epsilon$ limit taken first - which we take to
be $i\pi$\footnote{In this context, due to the $2i\pi$ periodicity of the
exponential, we need not worry about how the $x \rightarrow 0$ limit is
taken.} - and its value with the $\T$ limit taken first. 

If we take first the term $l=i$ then, looking at the
definition~\bref{eq:kdef}, $t(x,\T)$ terms of the first type are only
present if $M_{ij}(q(\pi k),\qbar(\pi k)$ is of the form $q(\pi
|k|)^{-2}\qbar(\pi |k|)^{-t_{i}-1} +$(terms in more negative powers of $q
,\qbar$). Expanding out equation~\bref{eq:mform}, we find $M$ is of the
form $q^{-x}\qbar^{-y} +$ more negative powers for given $x,y$, meaning we
are searching for the presence of a block $\{2,t_{i}+1\}$ in $S_{ij}(\T)$.
This is no surprise, as this is the block responsible for the pole
$\T_{h}+t_{i}\T_{H}$ discussed above; this is just another way of
representing the same situation. Case-by-case analysis shows this block is
present iff $G_{ij}$ is odd. 

Looking instead at $l \neq i$, we note that then
$[K_{il}]_{q\qbar}=-[G_{il}]_{\qbar}$ which, expanding out the definition
of the bracket notation, is $-\qbar^{G_{il}-1}-\qbar^{G_{il}-3} - \ldots -
\qbar^{-(G_{il}-1)}$. Thus, the only way a $\qbar'$ dependence can enter
is if $G_{il} > 1$. In this case, by an argument similar to the above, we
are now searching for $\{1,t_{i}\}$ blocks, with $t_{i} >1$.  Proceeding
again by case-by-case analysis, we find that the block $\{1,2\}$ is never
present and $\{1,3\}$ is only possible in $(g_{2}^{(1)},d_{4}^{(3)})$,
where it does not occur in the right element of the S-matrix to invoke
this process. 

To sum up, we find a single contribution to $y$ for $G_{ij}$ odd and
none otherwise. This can be restated as $y=-2i\pi \Theta(\T)G_{ij}$, showing
that we have indeed found a generalisation of the RTV formula. 

To complete this section, we must discuss the exceptional case
$a_{2n}^{(2)}$. Being self-dual, the S-matrix for this theory cannot be
found from the above. Following Oota, however, we note that the necessary
prescription is to replace each reference to $\rv\hv$ by $\hv=h=2n+1$,
take all $t_{a}=1$, and replace the incidence matrix by the ``generalised
incidence matrix''~\cite{Klassen}, which is obtained from the incidence
matrix of $a_{n}^{(1)}$ by replacing the last zero on the diagonal by
a one. Doing this, we obtain the correct integral S-matrix, and hence a
generalised RTV identity, for this case. 

\se{Multi-linear Identities}
\label{sec:multi}

The RTV result and its generalisation allow us to perform a simple trick
and generate a large number of S-matrix identities. Interchanging $i$ and
$j$ in~\bref{eq:grtv} does not change the lhs if $t_{i}=t_{j}$ - the two
roots are the same length - due to the symmetry of the S-matrix, so we can
equate the rhs before and after interchanging to get
\be
\prod_{l=1}^{r}\prod_{n=1}^{G_{il}}S_{jl}(\T+(2n-1-G_{il})\T_{H})=
\prod_{l'=1}^{r}\prod_{n'=1}^{G_{jl'}}S_{il'}(\T+(2n'-1-G_{jl'})\T_{H}).
\label{eq:multi}
\ee
(Note that the presence or absence of an exponential factor does not
affect this, as $t_{i}=t_{j}$ ensures $G_{ij}=G_{ji}$.) If $i$ and $j$ are
such that the corresponding rows of the incidence matrix consist of
entries no greater than 1, this reduces to
\be
\prod_{l=1}^{r}S_{il}(\T)^{G_{lj}}=\prod_{l'=1}^{r}S_{jl'}(\T)^{G_{l'i}},
\ee
and we can obtain identities for products of S-matrix elements, all
evaluated at the same rapidity; the first example,~\bref{eq:kexamp}, is
one of this set. Now, however, we also have identities in which not all
rapidities are equal. 

To generalise the connection between S-matrix product identities and
conserved charge sum rules to this case, we can take logarithmic
derivatives to find that if
\be
\prod_{a,b \in \{i,j\}}S_{ab}(\T+if^{1}_{ab})=\prod_{a',b' \in \{i',j'\}}
S_{a'b'}(\T+if^{2}_{a'b'}),
\ee
for some sets $\{i,j\}$ and $\{i',j'\}$ then
\be
\sum_{a,b \in \{i,j\}}\e^{-if_{ab}^{1}s}q_{s}^{a}q_{s}^{b}
=\sum_{a',b' \in \{i',j'\}} \e^{-if_{a'b'}^{2}s}q_{s}^{a'}q_{s}^{b'}.
\ee
Applying this to~\bref{eq:multi} gives
\be
\sum_{l=1}^{r}[G_{il}]_{\qbar(i\pi s)}q_{s}^{l}q_{s}^{j}=\sum_{l'=1}^{r}
[G_{jl'}]_{\qbar(i\pi s)}q_{s}^{l'}q_{s}^{i},
\label{eq:ccid}
\ee
where it should be noted that the sums over $n$ and $n'$
in~\bref{eq:multi} have been absorbed by the introduction of the
$[G_{ab}]_{\qbar(\pi s)}$ notation. 

To give a simple example of this result, in the $b_{r}^{(1)}$
algebra we have, for $1<i<r-1$
\be
S_{(r-1)(i-1)}(\T)S_{(r-1)(i+1)}(\T)=S_{i(r-2)}(\T)S_{ir}(\T+\T_{H})
S_{ir}(\T-\T_{H}),
\ee
and
\be
q_{s}^{r-1}q_{s}^{i-1}+q_{s}^{r-1}q_{s}^{i+1}=q_{s}^{i}q_{s}^{r-2}+\frac{1}{2}
\cos \frac{B^{b_{r}^{(1)}}\pi s}{2\rv\hv} \cdot q_{s}^{i}q_{s}^{r},
\ee
with (through the duality transformation $B^{b_{r}^{(1)}} \rightarrow 
2-B^{a_{2r-1}^{(2)}}$) corresponding 
identities for $a_{2r-1}^{(2)}$. 

\se{Other Observations}
\label{sec:other}
In the simply-laced cases, the conserved charges can be simply
characterised as components of the eigenvectors of the Cartan
matrix~\cite{Klassen}. Taking the logarithmic derivative of the
formula~\bref{eq:grtv} and proceeding as before allows us to generalise
this to nonsimply-laced cases, as follows: 
\be
\sum_{l=1}^{r}[G_{il}]_{\qbar(i\pi s)}q_{s}^{l}=2\cos\left[\pi s
\left(\frac{2-B}{2h}+\frac{Bt_{i}}{2\rv\hv}\right)\right]q_{s}^{i}.
\label{eq:eigen}
\ee
Note now, however, that the $t_{i}$ in the cos term stops this from being
a proper eigenvalue equation in the nonsimply-laced cases. In simply-laced
cases, this reduces to the known eigenvector result, since $[n]_{q}=n$ for
$n=0,1$ (as all entries of the incidence matrix are in these cases), and
we have all $t_{i}=1$ and $h=\rv\hv$. Rearranging, this can also be stated
as
\be
\sum_{l=1}^{r}[K_{il}]_{q(i\pi s)\qbar(i\pi s)}q_{s}^{l}=0.
\ee

We can also find a relation between the matrix $M$ and the conserved
charges. Noting that the S-matrix expression \bref{eq:Oota} explicitly
contains the matrix $M$, we first take the logarithmic derivative, and
then note that the resulting integral can be re-expressed as a contour
integral over the upper half-plane. The only poles in this expression are
in the matrix $M$, so, before we can continue, we must find a formula for
this. If we recall that the S-matrix can be written in a product
form~\cite{Dorey2} as 
\be 
S_{ab}(\T)=\prod_{x=1}^{h}\prod_{y=1}^{\rv\hv} \{x,y\}^{m_{ab}(x,y)} 
\ee
 (where the $\{x,y\}$ are standard building blocks, and the $m_{ab}(x,y)$s
are integers), and compare with Oota's integral form, we can find a
formula for $M$ as
\be
M_{ab}(q,\qbar)=\sum_{x=1}^{h}\sum_{y=1}^{\rv\hv}m_{ab}(x,y)\frac{q^{h-x}
\qbar^{\rv\hv-y}-q^{-(h-x)}\qbar^{-(\rv\hv-y)}}{q^{h}\qbar^{\rv\hv}-q^{-h}\qbar^{-\rv\hv}}. 
\label{eq:mform}
\ee 
This shows that the only poles present are at $k=im$, $m$ being any
integer, so the result is that we can re-express the integral in the form
of a Fourier expansion, and thus read off a relation between
$\varphi_{ab}^{(s)}$ and $M$ as
\be 
\varphi_{ab}^{(s)}=2 \sin \pi s \cdot \sinh s\T_{h} \cdot \sinh s\T_{H}
\cdot M(q(i\pi s),\qbar(i\pi s)).  
\ee
Of course, to find an expression in $q_{s}^{a}q_{s}^{b}$, we need to
include a scaling factor. Noting that
$\sum_{i=1}^{r}q_{s_{i}}^{a}q_{s_{i}}^{b}=\delta_{ab}$, where $s_{i}$ is
the $i$th exponent of a rank-$r$ algebra, we could use
$q_{s}^{a}q_{s}^{b}=\varphi_{ab}^{(s)}/\sum_{i=1}^{r}\varphi_{11}^{(s_{i})}$. 

Combining this with the expression for $M$, we get
\be
\varphi_{ab}^{(s)}=2 \sinh s\T_{h} \cdot \sinh s\T_{H} \cdot
\sum_{x=1}^{h}\sum_{y=1}^{\rv\hv}m_{ab}(x,y)\sin \left(
\frac{s\pi}{2}\left[ \frac{(2-B)x}{h}+\frac{By}{\rv\hv}\right]\right).
\ee
{}From this, it is straightforward to see that the matrix
$\varphi_{ab}^{(s)}$ is non-zero for generic $B$ by simple case-by-case
analysis. (This is different from this minimal case where, as noted by
Klassen and Melzer~\cite{Klassen}, we can get a zero matrix for
$s=\frac{h}{2}$ in simply-laced cases, even if that exponent is present.)
Had there been cases where $\varphi^{(s)}$ was zero for some $s$, then
taking the logarithmic derivative of an S-matrix identity would sometimes
have resulted in a trivial conserved charge identity. As it is, however,
we can always derive a non-trivial conserved charge identity from an
S-matrix identity and vice versa. 

\se{Conclusions}
We have found a generalisation of the RTV identity to nonsimply-laced
cases, and, from this, a way of generating S-matrix identities of the type
recently discussed by Khastgir. It is still an open question as to whether
we have found all such identities, or merely a subset. In addition, we
note that we can always use the technique of taking logarithmic
derivatives to generate corresponding identities for the conserved
charges. We have also, in equation~\bref{eq:eigen}, generalised the
``eigenvector'' characterisation of the conserved charges in the ATFTs
to the nonsimply-laced cases.

While this note was in preparation, the result~\bref{eq:grtv} was also
reported by Fring, Korff and Schulz~\cite{Fring} as their ``combined
bootstrap'' identity (we have altered our notation to accord with
theirs).  They proceeded by a geometrical argument, and went on to use
this to derive the integral formula for the S-matrix. They chose to
take the $\T$ limit last, and so their bootstrap identity does not
have the factor $\e^{-2i\pi G_{ij}\Theta(\T)}$, but it is otherwise
the same. The result that $S_{ij}(\T)$ contains the block
$\{2,t_{i}+1\}$ iff $G_{ij}$ is odd, which we found case-by-case, can 
also be found through a more systematic framework given by them.

\textbf{Acknowledgements:} we gratefully thank P. Dorey and R. Tateo for
discussions and many useful insights, as well as Andreas
Fring for pointing out a mistake in an earlier version of this
paper. We are also indebted to the United
Kingdom EPSRC for funding a PhD studentship. The work was partially
supported by a TMR grant from the European Community, contract number
ERBFMRXCT960012. 


\end{document}